\documentclass[aps,prd,nofootinbib]{revtex4-2}

\usepackage{physics}
\usepackage{color}
\usepackage{graphicx}
\usepackage{hyperref}
\hypersetup{colorlinks=true, citecolor=blue, urlcolor=blue, linkcolor=blue}

\begin{document}

\title{Twisted kink crystal in holographic superconductor}
\author{Masataka Matsumoto}
\author{Ryosuke Yoshii}
\affiliation{Department of Mathematics, Shanghai University, Shanghai 200444, China}
\affiliation{Center for Liberal Arts and Sciences, Sanyo-Onoda City University, Yamaguchi 756-0884, Japan}
\email{matsumoto@shu.edu.cn, ryoshii@rs.socu.ac.jp}
\begin{abstract}
Holographic superconductor model represents various inhomogeneous solutions with homogeneous sources. In this paper, we study inhomogeneous structures in the presence of the homogeneous current and the chemical potential. 
We find single complex kink condensates, multiple complex kinks condensates and twisted kink crystal condensates in which both the amplitude and the phase of the order parameter modulate in space. 
Analyzing the gauge-invariant phase difference in the single complex kink condensates, we find the non-monotonic behaviour with respect to the current. 
We confirm that the multiple complex kinks condensates and the twisted kink crystal condensates are well described by the analytic solutions obtained from the Gross-Neveu model or the Nambu-Jona-Lasinio model.
We also analyze the thermodynamic stability of the complex kink(s) condensates by computing the free energy.
Our results imply that holographic superconductor model provides the boundary physics which is effectively represented by the Ginzburg-Landau theory with higher corrections.
\end{abstract}

\maketitle

\section{Introduction}
Over the past two decades, strongly coupled systems have been investigated in the
framework of the Gauge/Gravity correspondence\,\cite{Maldacena1997,Gubser1998,Witten1998}.
The correspondence represents the duality between the strong coupling limit of the gauge theory and the weak coupling limit of the gravity theory.
The tractability of calculations in the gravity side enables one to analyze the non-perturbative properties in the strongly coupled system.
The framework has been prompted to deal with the non-equilibrium system\,\cite{Hubeny2010,Liu2018}.

Later the correspondence was also applied to condensed matter systems.
Along the line of such research, holographic superconductivity has been proposed to describe the superconducting phenomena in the context of the Gauge/Gravity correspondence\,\cite{Gubser2008,Hartnoll2008,Hartnoll2008k}.
It was successful in reproducing well-known properties of superconductivity though the setup is quite simple:\,it consists of the complex scalar fields coupled to Maxwell fields in the anti-de Sitte\,(AdS) background.
In spite of the evidences, it is still unclear how much the holographic superconductor can capture the superconductivity.
In condensed matter physics, there are two major frameworks to analyze the superconducting phenomena.
One is the Bardeen--Cooper--Schrieffer\,(BCS) model and the similar model, also known as the Nambu--Jona-Lasinio\,(NJL) model or the Gross--Neveu\,(GN) model used in different contexts (see \cite{Yoshii2019} and references therein).
The other is the Ginzburg--Landau\,(GL) model, which has been proposed as an effective model describing superconductors at the vicinity of the transition point and has been found to be obtained from the BCS model with proper approximations.
Both models are known to yield the basic properties of superconductivity.
Though it has been found that inhomogeneous solutions, such as a kink solution, exist in both theories, the usual GL model with second order derivatives, the mass term, and the quartic potential term do not have some solutions appearing in the BCS model.
Thus it can be the testbed of the holographic superconductors to check if the model corresponds to the GL model or the BCS model.

The inhomogeneous solutions naturally appear in the presence of inhomogeneous potentials and those cases have already been investigated in the context of holographic superconductors\,\cite{Flauger2010,Erdmenger2013}.
Interestingly, the existence of the inhomogeneous solutions in the absence of the inhomogeneity of the system has also been shown\,\cite{Keranen2009,Keranen22009,Lan2017}.
It is known that this type of inhomogeneous solutions are realized in terms of the condensed matter physics.
For example, the stabilization of the Larkin--Ovchinnikov--Fulde--Ferrel\,(LOFF) phase, in which the phase and/or the amplitude of the order parameter modulates in space, by the homogeneous magnetic fields are shown\,\cite{LO,FF}.
The experimental evidence of the LOFF phase have also been shown in a heavy fermion compound system\,\cite{Radovan2003}, an organic superconductor\,\cite{Yonezawa2008}, and in the presence of the ultra-cold atomic gases\,\cite{Liao2010}.

In our previous work\,\cite{Matsumoto2019}, we partially answered to the above question.
We have shown that the holographic superconductor model reproduces various inhomogeneous solutions for the homogeneous setup.
We have also shown that inhomogeneous solutions, which are not included in the standard GL model, appear in the holographic superconductors and the qualitative consistency with inhomogeneous solutions in the BCS model has been checked.
It implies that the holographic superconductors describe the superconducting phenomena beyond the GL theory with second order derivatives, the mass term, and the quartic potential term.
However, our previous study is restricted to the system without a current.
In the presence of a current, it is expected that the order parameter becomes complex and thus the richer phase structure appears. 
Thus, a further check of the applicability of the holographic superconductors should be carried out and it might shed light on the usage of holographic superconductors for nonequilibrium system.

In this paper, we extend the method employed in oure previous paper to analyze the superconducting system with a current.
In the holographic superconductor, the homogeneous condensate with a current was studied in\,\cite{Herzog2008,Arean2010} and the inhomogeneous setup for studying the holographic Josephson junction was presented in\,\cite{Horowitz2011}. Also, the inhomogeneous condensate in which a kink structure moves with a constant velocity was studied in\,\cite{Gao2019}. The holographic vortices and vortex-lattice have been studied in non-perturbative approaches\,\cite{Montull2009,Keranen32009,Xia2019,Donos2020}. However, the one-dimensional modulated solutions with a constant current, such as complex kink(s) and twisted kink crystals, are poorly studied.
Once the (metastable) inhomogeneous solutions in the presence of the current are found and shown to be consistent with the previously known solutions (such as complex kink(s) solutions or twisted kink crystal solutions) it will be more evident that the holographic superconductors describe the superconducting phenomena.
Moreover, if the solutions are not included in the standard GL with the second-order derivatives, the mass term, and the quartic potential term, it implies that the holographic superconductors capture the superconductivity beyond the standard GL level. To ensure this, we analyze the quantitative coincidence with the GL theory using higher corrections.

This paper is organized as follows.
In section \ref{sec:2}, we review the holographic setup in our study.
In order to find spatially inhomogeneous solutions, we will derive the equations of motion by assuming that each field depends on the spatial coordinate in addition to the AdS radial coordinate.
In section \ref{sec:3}, we numerically solve these equations and find solutions which have an inhomogeneous amplitude and phase of the condensate.
We show the single complex kink condensate, the multiple complex kinks condensate, and the twisted kink crystal condensate as inhomogeneous solutions to the equations. 
We also study the relation between our solutions and the known analytic solutions derived from the GN model and the NJL model.
In addition, we compute the free energy of those solutions.
Section \ref{sec:4} is devoted to conclusion and discussion.

\section{Holographic setup} \label{sec:2}
In this section, we present the holographic superconductor model at finite temperature.
A key tenet of the holographic superconductor is mapping the strongly correlated system to the dual theory in which the scalar field is coupled to the background electromagnetic field in the curved spacetime.
In this paper, we consider the Einstein-Maxwell theory coupled to a charged complex scalar field with a negative cosmological constant in (3+1) dimensional spacetime\,\cite{Gubser2008,Hartnoll2008,Hartnoll2008k}.
In this setup, the dual boundary theory is in (2+1) dimensional spacetime.
In our study, we take the probe approximation in which we ignore the backreaction of the gauge and the scalar fields on the metric.
In the probe approximation, the background geometry is independent of the matter sector.
We consider a (3+1) dimensional planar AdS black hole as the background metric:
\begin{equation}
    ds^{2}=\frac{L^{2}}{z^{2}} \left[ -f(z) dt^{2} + \frac{dz^{2}}{f(z)} + dx^{2}+dy^{2} \right],
    \label{eq:metric}
\end{equation}
where
\begin{equation}
    f(z)=1-\left( \frac{z}{z_{H}} \right)^{3}.
\end{equation}
Here, $z_{H}$ is the black hole horizon and the AdS radius is given by $L$.
The Hawking temperature, which corresponds to the heat-bath temperature of dual field theory, is given by $T=3/(4 \pi z_{H})$.
We assume that the system consists of the complex scalar field $\Psi$ interacting with a $U(1)$ gauge field $A_\mu$.
The dynamics of the system is determined by the following action
\begin{equation}
    S=\int d^{4}x \sqrt{-g} \left( -\frac{1}{4} F_{\mu\nu}F^{\mu\nu}-\left| D_{\mu} \Psi \right|^{2} +V\left( \left| \Psi \right| \right) \right),
    \label{eq:action}
\end{equation}
where the covariant derivative is defined by $D_{\mu}\Psi=\left( \partial_{\mu} -iA_{\mu} \right)\Psi$. 
Here the field strength is given by $F_{\mu\nu}=\partial_{\mu}A_{\nu}-\partial_{\nu}A_{\mu}$ and $g=\det g_{\mu\nu}$.
We assume that the potential is given by $V=-m^{2} \left|\Psi \right|^{2}$. In this paper, we will take the scalar mass to be $m^2=-2/L^2$ which is above the Breitenlohnor-Friedman bound\,\cite{BF}.
For simplicity, we set $L=1$. 
The equation of motion for the complex scalar field is
\begin{eqnarray}
    0=\frac{1}{\sqrt{-g}} D_{\mu} \left( \sqrt{-g} D^{\mu}\Psi\right) -m^{2} \Psi.
\end{eqnarray}
The Maxwell equation is
\begin{eqnarray}
\frac{1}{\sqrt{-g}} \partial_{\mu} \left( \sqrt{-g} F^{\mu\nu}\right) =i (\Psi^{*} \partial^{\nu} \Psi-\Psi \partial^{\nu} \Psi^{*}) - 2A^{\nu}\left| \Psi \right|^{2}.
\end{eqnarray}

In order to study the inhomogeneous solutions in the presence of a current, the ansatz for our fields are given by
\begin{equation}
	\Psi(z,x)=\psi(z,x)e^{i \varphi(z,x)}, \hspace{1em} A=A_{t}(z,x)dt+A_{z}(z,x)dz+A_{x}(z,x)dx.
\end{equation}
Here, all variables $\psi$, $\varphi$, $A_{t}$, $A_{z}$, and $A_{x}$ are real functions of $z$ and $x$.
By using the gauge degrees of freedom, we define the gauge invariant fields by $M_{\mu}=A_{\mu}-\partial_{\mu}\varphi$.
Thus, we have four independent fields:\,$\psi$, $M_{t}$, $M_{z}$, and $M_{x}$.
For later convenience, we also define the scalar field $\psi=z \phi/\sqrt{2}$.
The equations of motion for those fields are given by\footnote{Note that the last term of\,(\ref{eq:eom1}) is derived from the term of $z^{-2}f^{-1}(z f'-L^2 m^2 - 2 f) \phi$.}
\begin{eqnarray}
	\phi''+\frac{\partial_{x}^{2}\phi}{f}+\frac{f'}{f}\phi'+\left(\frac{M_{t}^{2}}{f^{2}}-M_{z}^{2}-\frac{M_{x}^{2}}{f}-\frac{z}{z_{H}^3 f}\right)\phi &=&0, \label{eq:eom1}\\
	M_{t}''+\frac{\partial_{x}^{2}M_{t}}{f}-\frac{M_{t}\phi^{2}}{f}&=&0, \label{eq:eom2}\\
	\partial_{x}^{2}M_{z}-\partial_{x}M_{x}'-M_{z}\phi^{2}&=&0, \label{eq:eom3}\\
	M_{x}''-\partial_{x}M_{z}'+\frac{f'}{f}\left(M_{x}'-\partial_{x}M_{z}\right)-\frac{M_{x}\phi^{2}}{f}&=&0, \label{eq:eom4}\\
	M_{z}'+\frac{\partial_{x}M_{x}}{f}+\frac{2}{\phi}\left(M_{z}\phi'+\frac{M_{x}\partial_{x}\phi}{f}\right)+\frac{f'}{f}M_{z}&=&0, \label{eq:eom5}
\end{eqnarray}
where the prime denotes the derivative with respect to $z$.
Here, Eq.\,(\ref{eq:eom5}) is obtained from the conservation of the source in Maxwell equation.
Since Eq.\,(\ref{eq:eom5}) is not independent from the other equations, we have four equations of motion and four variables.

Now we consider the boundary conditions for these fields.
By analyzing the above equations of motion near the AdS boundary $z=0$, one can expand these fields as follows:
\begin{eqnarray}
	\phi &=& \phi^{(1)}+\phi^{(2)}z+{\cal O}(z^{2}), \label{eq:asym1}\\
	M_{t} &=& \mu+\rho z+ {\cal O}(z^{2}),  \label{eq:asym2}\\
	M_{z} &=& {\cal O}(z),  \label{eq:asym3}\\
	M_{x} &=& \nu +J z+ {\cal O}(z^{2}).  \label{eq:asym4}
\end{eqnarray}
Here $\mu$, $\rho$, $\nu$, and $J$ are interpreted as the chemical potential, the charge density, the velocity, and the current in the boundary field theory, respectively.\footnote{In the asymptotic form of $M_z$ near the AdS boundary, we impose $2 \phi^2 M_z - \partial^2_{x} M_{z} +\partial_{x}M'_{x} =0$ from the zeroth order of $z$. If we assume $M_{z}(z=0,\rho)=0$, the spatial homogeneity of the current is clearly satisfied since $\partial_{x}M'_{x}=\partial_{x}J=0$ at the AdS boundary.}
In our study, we assume that $\phi^{(1)}=0$ and $\phi^{(2)}=\expval{O_{2}}(x)$, which is the order parameter given as the function of $x$.
In addition, we assume that the chemical potential $\mu$ and the current $J$ are homogeneous in the spatial coordinate $x$.
The latter assumption is consistent with the conservation of the current.

At the horizon $z=z_{H}$, we impose the regularity conditions for variables as the boundary conditions.
We choose the condition of $M_{t}$  as $M_{t}=0$ at the horizon.
This artificial gauge choice has been discussed in\,\cite{Gubser2008}.
The other boundary conditions are explicitly written as
\begin{eqnarray}
	\phi'-\frac{1}{3}\partial_{x}^{2}\phi+\frac{1}{3}\left(1+M_{x}^{2}\right)\phi &=&0, \\
	M_{x}'-\partial_{x}M_{z}+\frac{1}{3}M_{x}\phi^{2} &=&0, \\
	M_{z}-\frac{1}{3}\partial_{x}M_{x}-\frac{2M_{x} \partial_{x}\phi}{3\phi}&=&0.
\end{eqnarray}
In addition to these boundary conditions, we require the conditions at $x = \pm\infty$.
At these points, we require that all the functions approach the homogeneous solutions.
In other words, we need to impose the Neumann boundary conditions for all variables at $x = \pm\infty$.
For numerical convenience, however, we impose the boundary conditions at $x=0$ and $x=\infty$ by using the fact that $\phi$, $M_t$, and $M_x$ are even and $M_z$ is an odd function of $x$. Therefore, we impose the Neumann boundary condition for $\phi$, $M_t$, and $M_x$, and the Dirichlet boundary condition for $M_z$ at $x=0$.
For technical reasons, we set the calculation region to $-l/2 \leq x\leq l/2$.
Then, we confirm that the solution asymptotes to a fixed profile with a sufficiently large $l$.

The gauge-invariant phase difference is defined by $\gamma \equiv \Delta \varphi - \int A_{x}$.
In our ansatz, it can be explicitly written by
\begin{equation}
	\gamma= -\int dx \left[ \nu(x)-\nu(\pm\infty) \right].
	\label{eq:phase}
\end{equation}
The second term of the integrand in\,(\ref{eq:phase}) is just a regulator since the homogeneous solutions at $x=\pm\infty$ have a constant current and a constant velocity.

It should be noted that equations of motion are invariant under the following scaling symmetry:
\begin{equation}
	(t,z,x,z) \to \lambda (t,z,x,y), \hspace{1em} (\phi,M_{t},M_{z},M_{x}) \to \frac{1}{\lambda} (\phi,M_{t},M_{z},M_{x}),
\end{equation}
where $\lambda$ is a rescaling parameter.
We fix $z_{H}=1$ by using this scaling symmetry.
Thus, we choose $T/\mu$ and $J/\mu^{2}$ as the scaled parameters which characterize the system.

Since the equations of motion are partial nonlinear differential equations, we solve them numerically. 
To perform the numerical calculation, we use the Chebyshev pseudospectral method. In our study, we use 21 points along the $z$ direction and 51 points along the $x$ direction. 
In order to obtain solutions, we prepare an appropriate initial configuration and employ the Newton-Raphson relaxation method. After iteration in the Newton-Raphson relaxation scheme, we confirm that the configurations for each field satisfy the equations of motion.
In the following calculations, we set $l=10$ as the calculation region.

\begin{figure}[htbp]
\centering
\includegraphics[width=7.5cm]{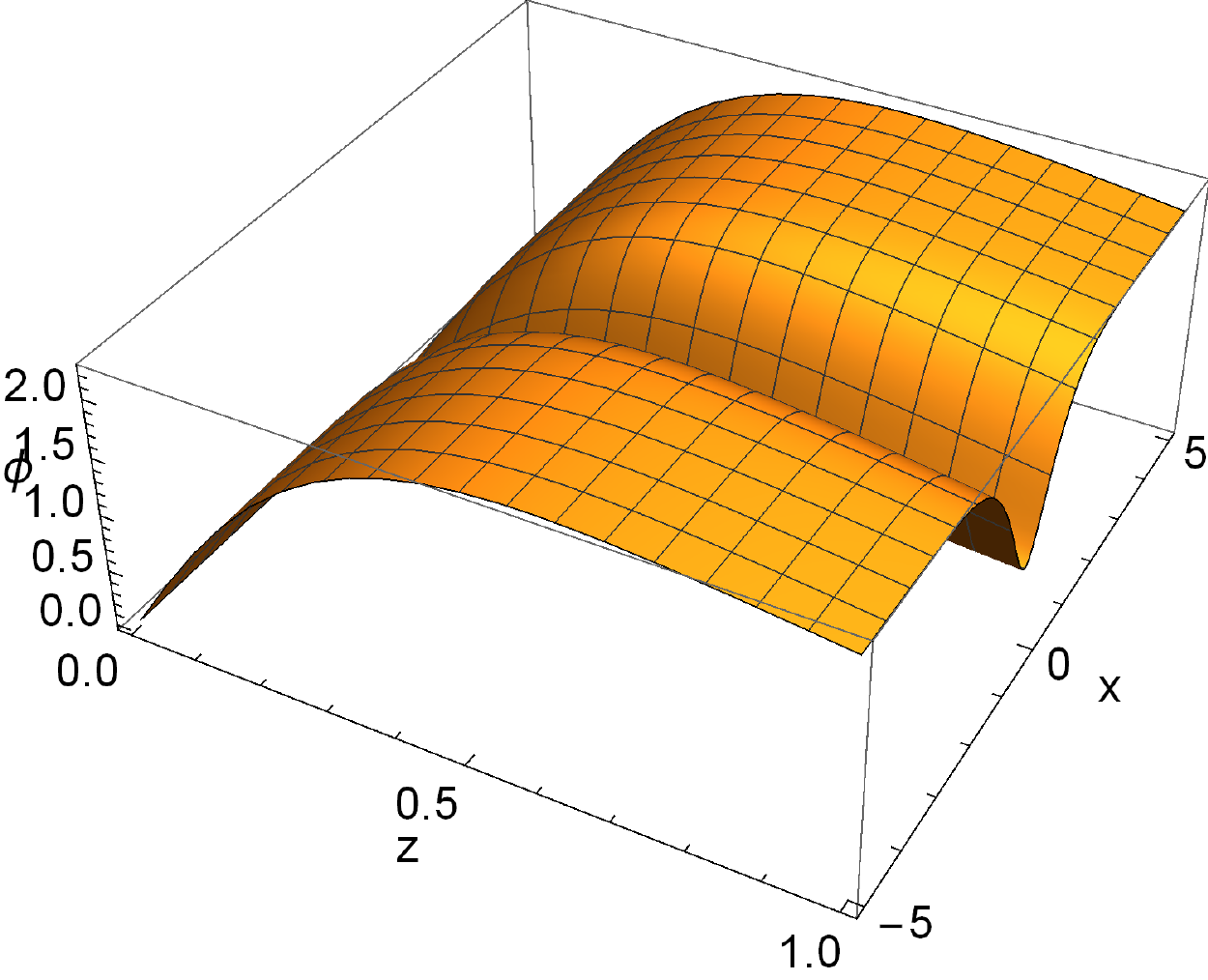}
\includegraphics[width=7.5cm]{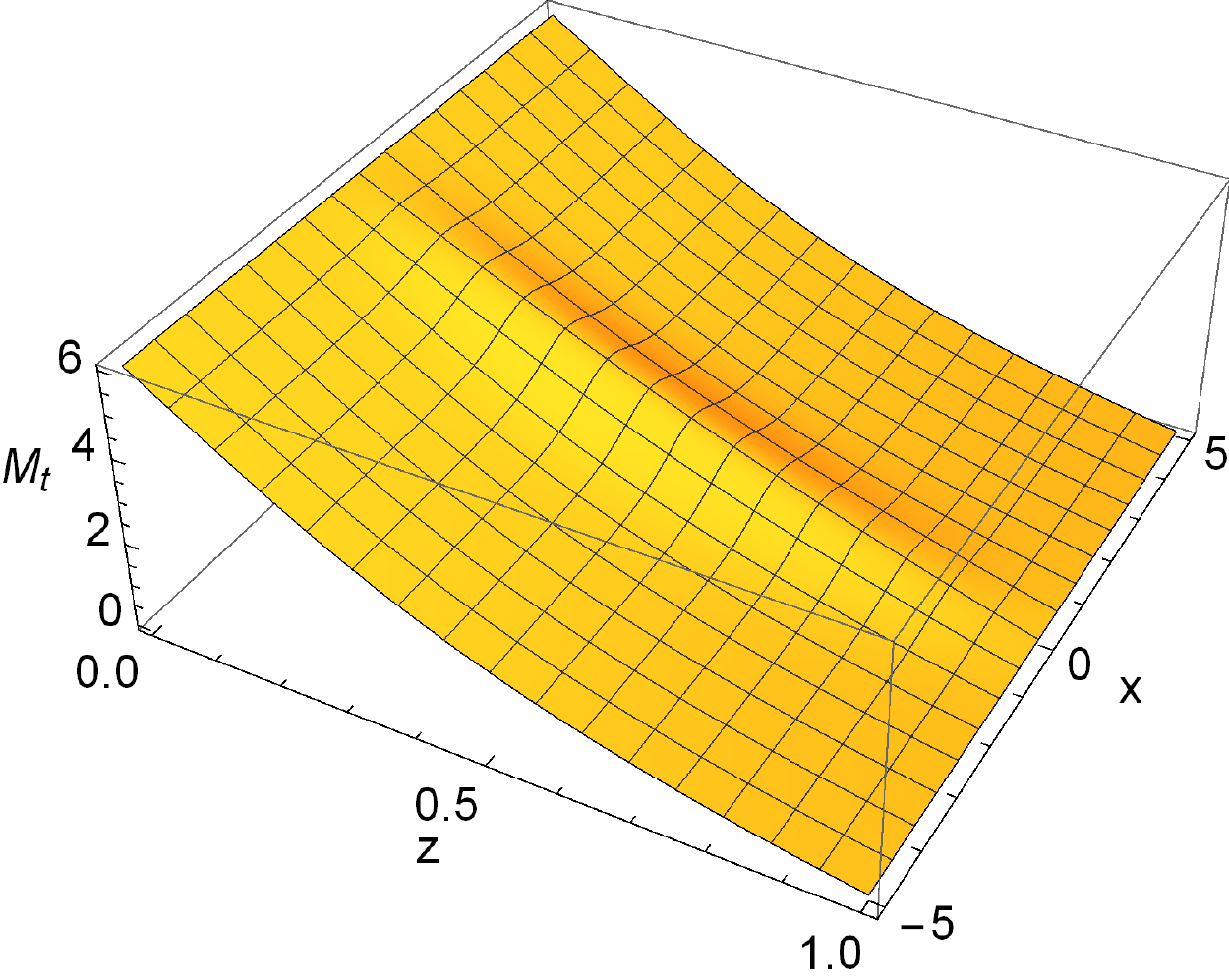}
\includegraphics[width=7.5cm]{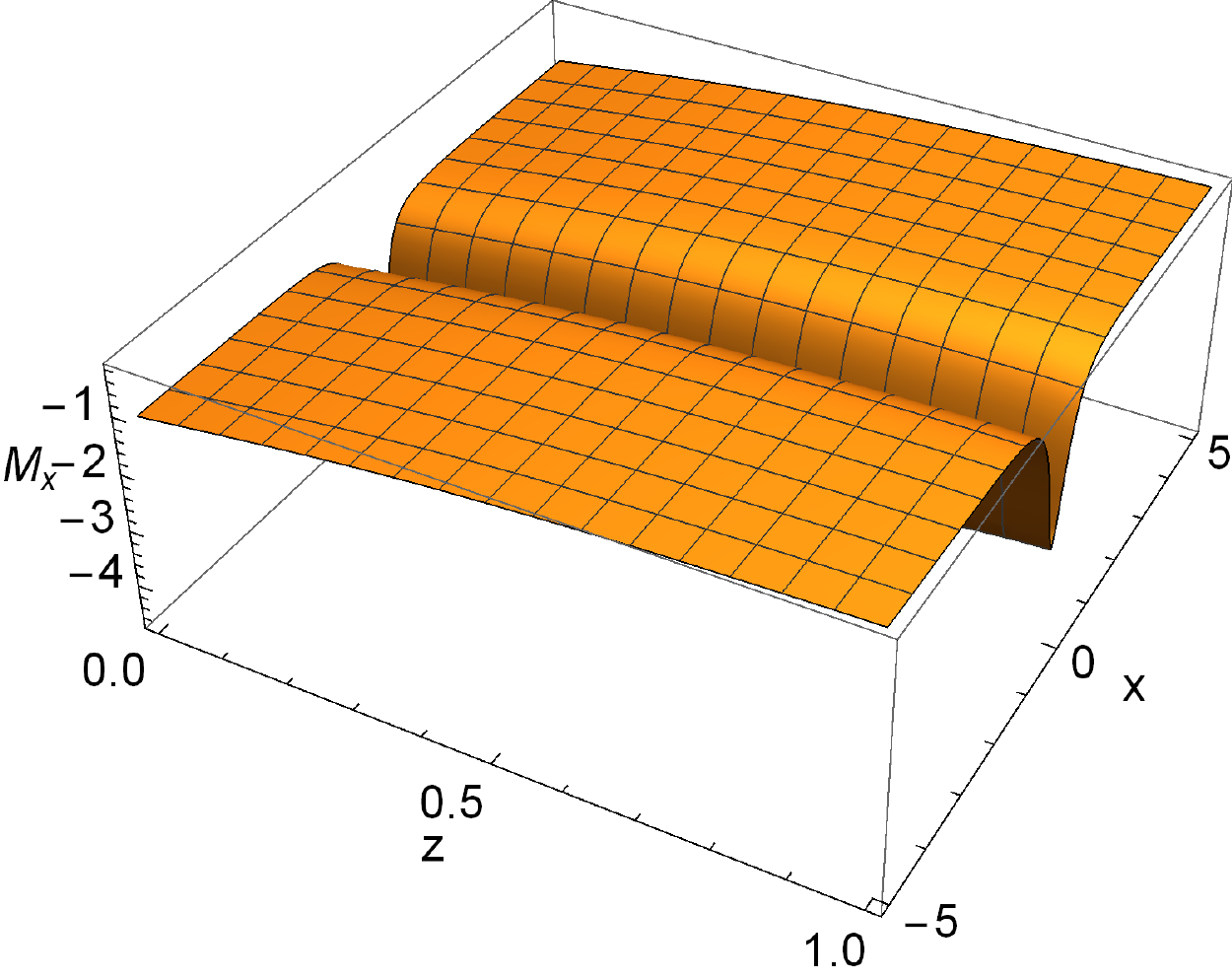}
\caption{The typical profiles of $\phi(z,x)$, $M_{t}(z,x)$, and $M_{x}(z,x)$ for $T/\mu=0.0398$ and $J/\mu^{2}=0.0417$.}
\label{fig:3Dplot}
\end{figure}

\section{Inhomogeneous solutions} \label{sec:3}
In this section, we show the inhomogeneous solutions, particularly the single complex kink condensate, multiple complex kink condensate, and twisted kink crystal condensate, by numerically solving the equations of motion\,(\ref{eq:eom1})-(\ref{eq:eom4}).

\subsection{Single complex kink}

\begin{figure}
	\centering
	\includegraphics[width=7.5cm]{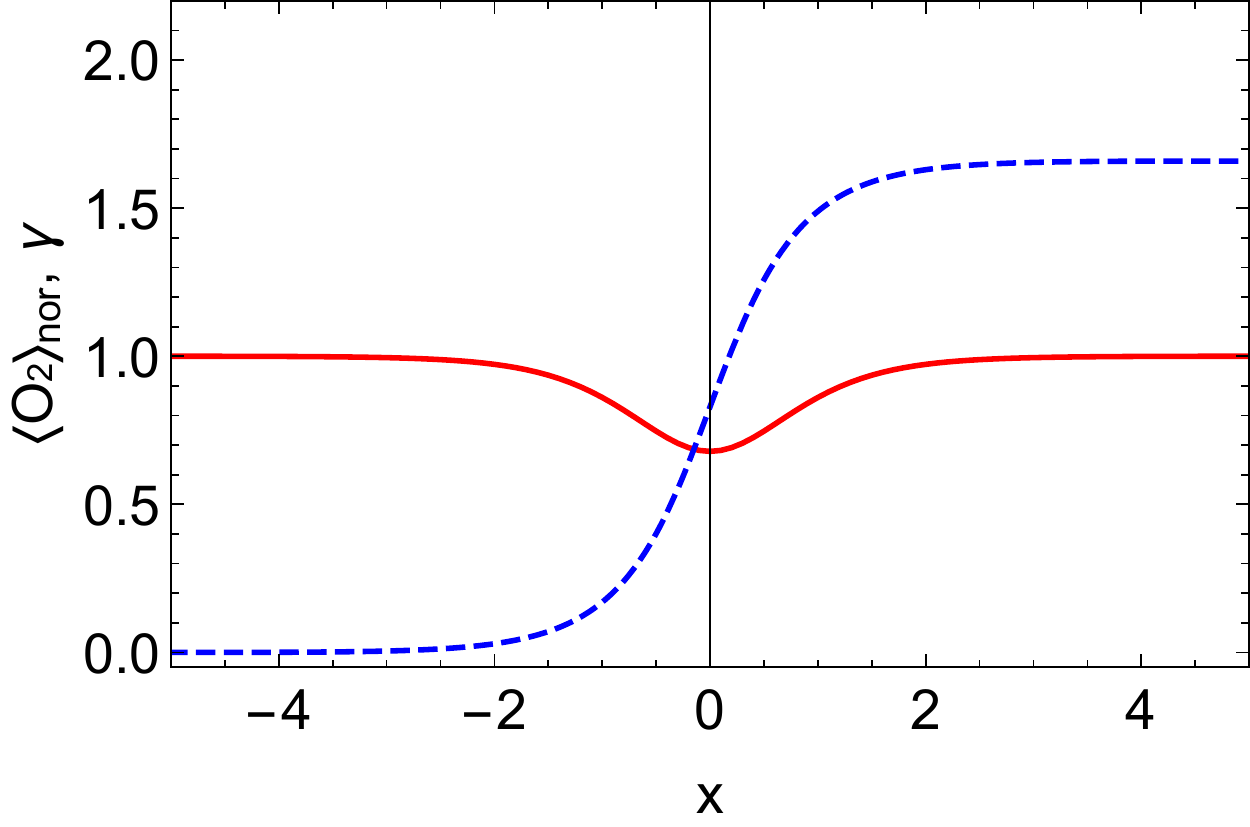}
	\includegraphics[width=7.5cm]{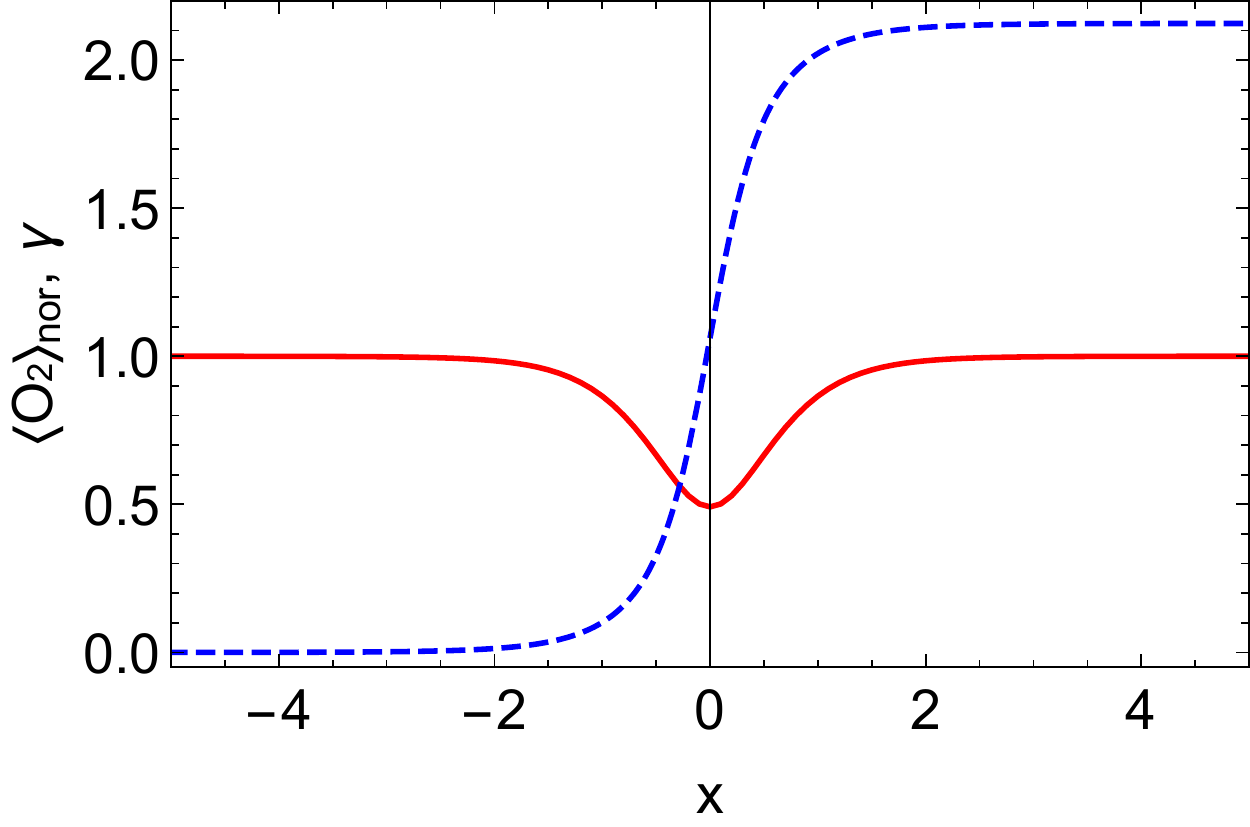}
	\caption{The normalized amplitude of the condensate $\expval{O_{2}}_{\rm nor}$\,(solid) and the gauge-invariant phase $\gamma$\,(dashed) as a function of $x$ for $J/\mu^{2}=0.0639$\,(left panel) and $J/\mu^{2}=0.0556$\,(right panel).}
	\label{fig:cond}
\end{figure}

As an example, we show the typical profiles of $\phi(z,x)$, $M_{t}(z,x)$, and $M_{x}(z,x)$ for $T/\mu =0.0398$ and $J/\mu^{2}=0.0417$ in Fig.\,\ref{fig:3Dplot}.
As can be seen from Fig.\,\ref{fig:3Dplot}, we find that each bulk field has an inhomogeneous structure.
We can compute the order parameter $\expval{O_{2}}$, the charge density $\rho$, and the velocity $\nu$ from each solution by using the asymptotic form Eqs.\,(\ref{eq:asym1}), (\ref{eq:asym2}), and (\ref{eq:asym4}).
Then, we obtain the amplitude and the phase of the order parameter as a function of $x$ as shown in Fig.\,\ref{fig:cond}.
Here, we introduce a normalized condensate which is defined by $\expval{O_{2}}_{\rm nor} \equiv \expval{O_{2}}/ \left.\expval{ O_{2}}\right|_{x=\pm\infty}$.
These results imply that the condensate gets close to the homogeneous solutions and the phase difference becomes smaller as $J/\mu^{2}$ increases.
It has been known that this type of inhomogeneous solution corresponds to the {\it single complex kink condensate} in which both the amplitude and the phase of the condensate have the spatially inhomogeneous profile as shown in Fig.\,\ref{fig:cond}\,\cite{Basar2008,Basar2008ki}.
As studied in\,\cite{Keranen2009}, the holographic superconductor model exhibits a single real kink condensate in the presence of the spatially constant chemical potential.
Compared to this type of solution, the amplitude of order parameter in the single complex kink condensate does not reach zero at the position of the kink\,($x=0$ in our case).
We also find that the phase of the order parameter suddenly changes at that position.
It is also found that the dip of the condensate becomes larger for larger phase difference.
This behaviour can be understood by the energetics; both the phase gradient and the amplitude modulation of the condensate give kinetic energy
and thus the (meta-)stable solutions are obtained by the solutions which minimize $|\Psi^\ast \nabla \Psi|^2$.
If we focus on the phase modulation, we note it should take place in the region where the magnitude of the condensate is smaller and the sudden change of the phase is not favored. 
On the other hand, if we focus on the magnitude of the condensate, we see that the sudden drop of the magnitude of condensate costs the energy.
This trade-off relation results in the behaviour shown in Fig.\,\ref{fig:cond}.
If the limit of the phase difference is $\pi$,
the solution becomes the ordinal kink solution in which the condensate becomes zero at the node.
The behaviour of our single complex kink condensate is qualitatively consistent with that found in the NJL model\,\cite{Shei1976,Basar2008ki}.

\begin{figure}[tbp]
	\centering
	\includegraphics[width=7.5cm]{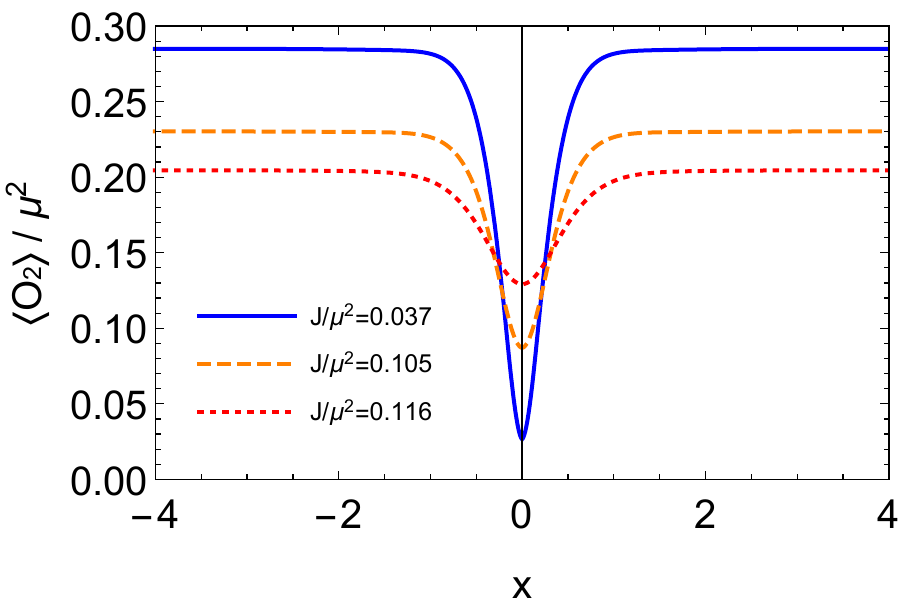}
	\includegraphics[width=7.5cm]{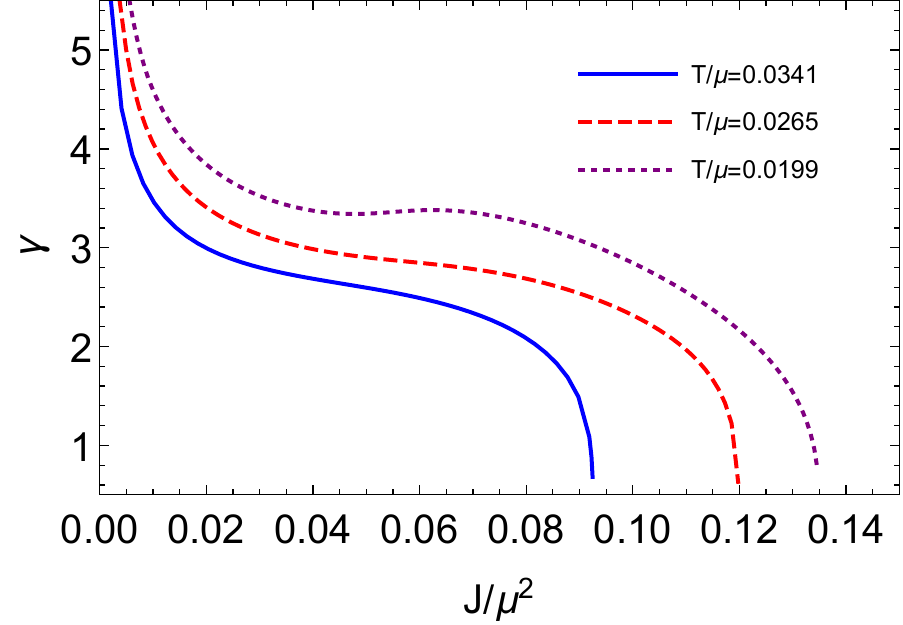}
	\caption{The left panel shows profiles of the condensate $\expval{O_{2}}_{\rm nor}$ for several values of $J/\mu^{2}$ with $T/\mu=0.0265$. The right panel shows the profile of the gauge-invariant phase difference $\gamma$ as a function of $J/\mu^{2}$ for $T/\mu=0.0341$\,(solid), $T/\mu=0.0265$\,(dashed), and $T/\mu=0.0199$\,(dotted).}
	\label{fig:cond2}
\end{figure}

Furthermore, we study the $J/\mu^{2}$ dependence of the single complex kink condensates with $T/\mu$ fixed.
In the left panel of Fig.\,\ref{fig:cond2}, we show the plots of condensate for various values of $J/\mu^{2}$ at a fixed value of $T/\mu$.
We also show the total gauge-invariant phase difference as a function of $J/\mu^{2}$ for several values of $T/\mu$ in the right panel of Fig.\,\ref{fig:cond2}.
For lower $J/\mu^{2}$, the condensate profile becomes sharp at the kink position and the total gauge-invariant phase difference rapidly changes.
It is technically difficult to apply our numerical method near $J/\mu^{2}\sim 0$ due to the lack of numerical accuracy.
However we expect that the total gauge-invariant phase difference approaches to $2\pi$ when $J/\mu^2$ reduces,
since the condensate with the phase winding of $2\pi$ corresponds to the homogeneous one.
For larger $J/\mu^{2}$, we find that the condensate approaches to the homogeneous profile and the total gauge-invariant phase difference goes to zero.
It appears that the value of $J/\mu^{2}$ at which the total gauge-invariant phase difference for the single complex kink condensate becomes zero coincides with the critical value of $J/\mu^{2}$ for the homogeneous condensate beyond which the normal state becomes the ground state\,\cite{Arean2010}.
Interestingly, we find that the total gauge-invariant phase difference shows a non-monotonic behaviour as a function of $J/\mu^{2}$ at temperatures much lower than the chemical potential.
For instance, the total-gauge invariant phase difference as a function of $J/\mu^{2}$ is a monotonically decreasing function for larger $T/\mu$ (solid line in the right figure of Fig.\,\ref{fig:cond2}).
Since the larger current tends to break the condensate, it is consistent that the inhomogeneous structure can be suppressed and the total gauge-invariant phase difference becomes small.
On the other hand, its behaviour changes to non-monotonic for smaller $T/\mu$ (dashed and dotted lines in the right figure of Fig.\,\ref{fig:cond2}).
This characteristic behaviour implies that there could be a phase transition at specific temperature.

\subsection{Multiple complex kinks}
\begin{figure}[tbp]
	\centering
	\includegraphics[width=7.5cm]{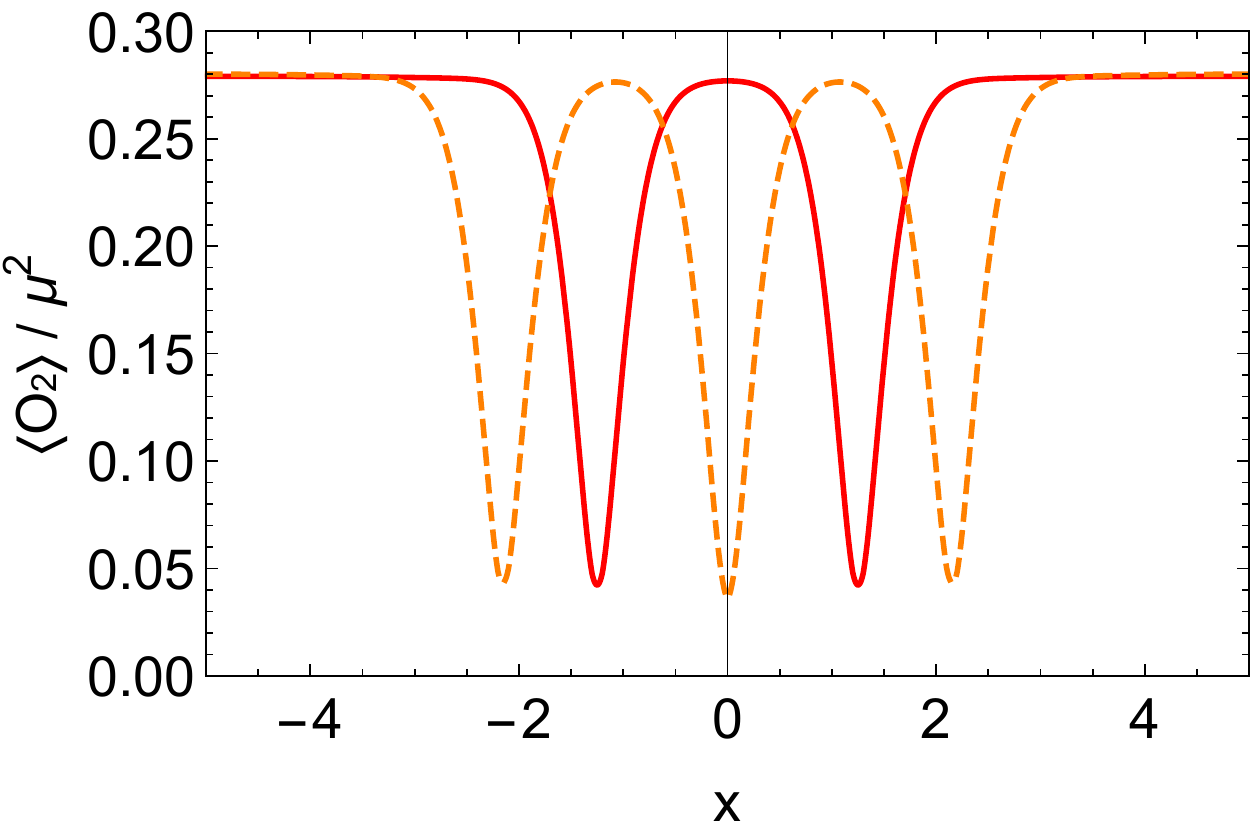}
	\includegraphics[width=7.5cm]{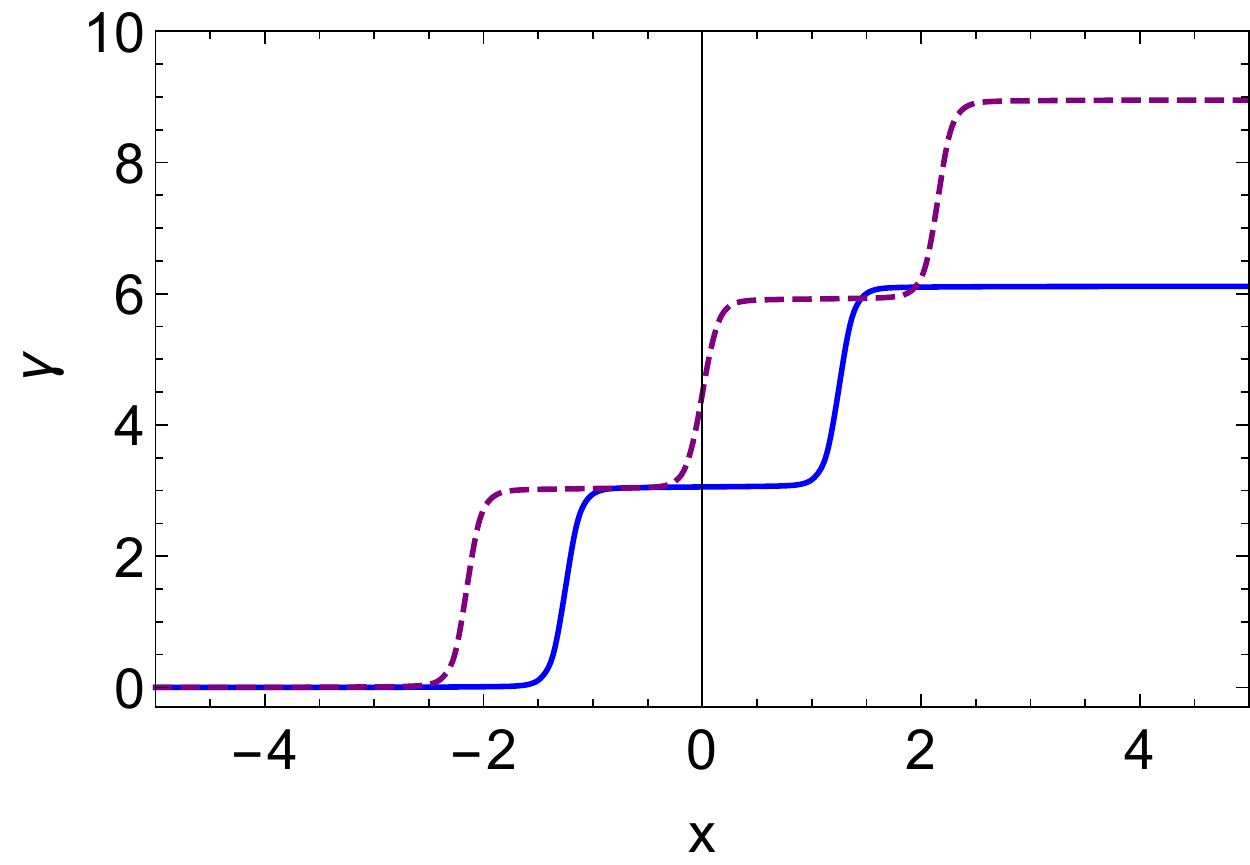}
	\caption{The plots show the amplitude\,(left panel) and the phase\,(right panel) of the order parameter for $T/\mu=0.0265$ and $J/\mu^{2}=0.0494$, respectively. The solid line denotes the two complex kinks condensate and the dashed line denotes the three complex kinks condensate.}
	\label{fig:multikink}
\end{figure}

In addition to the single complex kink condensate, we also find the solutions which have two or more kinks. We show the amplitude and the phase of the order parameter for this type of solution in Fig.\,\ref{fig:multikink}.
As can be seen from Fig.\,\ref{fig:multikink}, two or three kinks lie close to each other and their amplitudes approach to the homogeneous condensates at the boundary of the $x$ coordinate. Also, the phase of the order parameter suddenly changes at the position of the kinks, which is the same behaviors as that of the single complex kink condensate. Therefore, this type of solution corresponds to the {\it multiple complex kinks condensate}.
It is known that the multiple complex kinks condensate is obtained from the GN model and the NJL model or the GL theory with higher derivatives and with the higher order potential terms \cite{Takahashi2012, Takahashi2013}.
Thus, our result implies that the holographic superconductor model succeeds to describe the superconducting phenomena beyond the conventional GL model,\footnote{In \cite{Matsumoto2019}, we provided the same assertion by finding the multiple real kinks condensate in the absence of the current.}
which has only the second order derivatives, the mass term and the quartic potential term.
We quantitatively confirm this expectation later.

\subsection{Twisted kink crystal}
\begin{figure}[tbp]
	\centering
	\includegraphics[width=7.5cm]{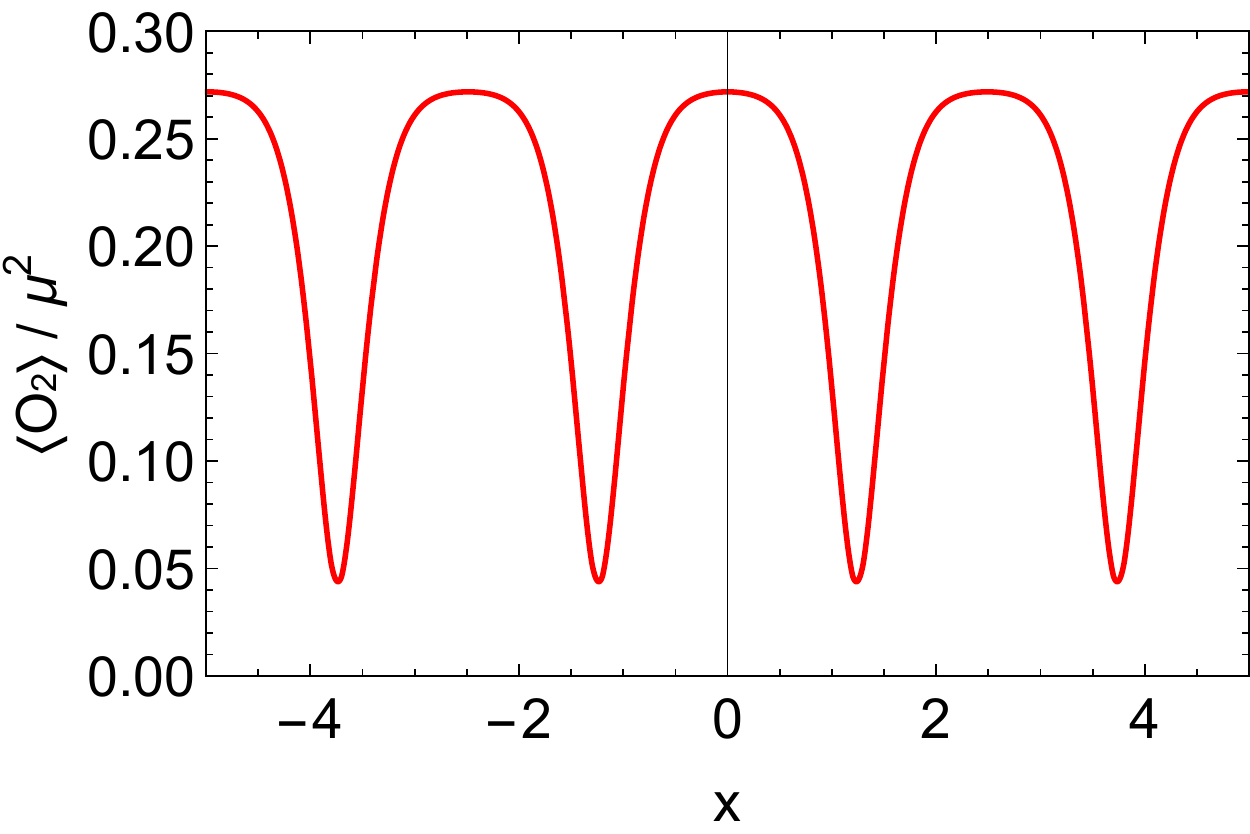}
	\includegraphics[width=7.5cm]{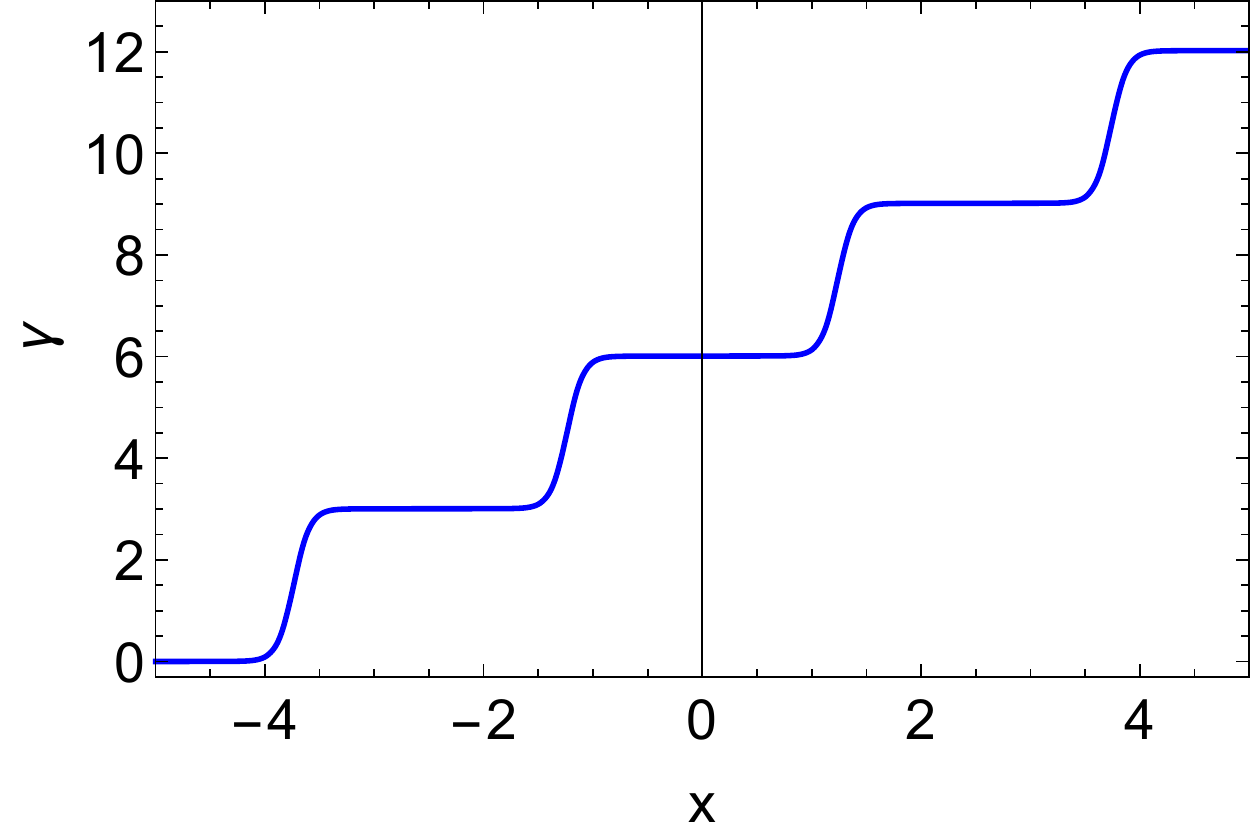}
	\caption{The plots show the amplitude\,(left panel) and the phase\,(right panel) of the order parameter in the twisted kink crystal condensate for $T/\mu=0.0265$ and $J/\mu^{2}=0.0617$, respectively.}
	\label{fig:TKC}
\end{figure}

Here, we also study the complex condensate in the periodic system.
We impose the periodic boundary condition for the fields at the spatial boundary $x=\pm l/2$, instead of the Neumann boundary condition as mentioned above. 
Fig.\,\ref{fig:TKC} shows a typical solution in which the amplitude and the phase of the order parameter periodically modulate in space.
This type of solution corresponds to the so-called {\it twisted kink crystal condensate}, which has been studied in terms of the GN model and the NJL model\,\cite{Basar2008,Basar2008ki}. It has been known that the twisted kink crystal condensate is also derived from GL theory with higher corrections such as higher derivative terms and higher order potential terms as shown in the following discussion.

\subsection{Boundary interpretation}

So far we have focused on what types of inhomogeneous condensates are obtained from the bulk solutions. 
In this section, we show the analytic forms of the inhomogeneous condensates in terms of the field theory and confirm if our solutions agree with them. 
In the (1+1) dimensional GN model or NJL model, those types of inhomogeneous solutions are written in the analytic forms. 
For instance, the two complex kinks solution is given by\,\cite{Takahashi2012,Takahashi2013}
\begin{equation}
	\Delta(x) = -i m + 2\left( e^{-i \theta_{1}} h_{1}(x) + e^{-i \theta_{2}} h_{2}(x) \right),
	\label{eq:2kink}
\end{equation}
where
\begin{eqnarray}
	h_{1}(x) &=& \frac{-\kappa_{1}\left(1+e^{-2\kappa_{2}(x-x_{2})} \right)+\alpha \sqrt{\kappa_{1} \kappa_{2}}}{\left( 1+e^{-2\kappa_{1}(x-x_{1})}\right) \left( 1+e^{-2\kappa_{2}(x-x_{2})} \right) - \left| \alpha \right|^{2}}, \\
	h_{2}(x) &=& \frac{-\kappa_{2}\left(1+e^{-2\kappa_{1}(x-x_{1})} \right)+\alpha^{*} \sqrt{\kappa_{1} \kappa_{2}}}{\left( 1+e^{-2\kappa_{1}(x-x_{1})}\right) \left( 1+e^{-2\kappa_{2}(x-x_{2})} \right) - \left| \alpha \right|^{2}}.
\end{eqnarray}
Here, $\kappa_{1}$, $\kappa_{2}$, and $\alpha$ are written as functions of $m$, $\theta_{1}$, and $\theta_{2}$:
\begin{equation}
	\kappa_{1}= m\sin\theta_{1}, \hspace{1em} \kappa_{2}=m\sin \theta_{2}, \hspace{1em} \alpha=\frac{2\sqrt{\kappa_{1}\kappa_{2}}}{im\left(e^{-i\theta_{1}}-e^{i\theta_{2}} \right)}.
\end{equation}
Since $m$ is an overall factor, this solution is determined by four parameters: $\theta_{1}$, $\theta_{2}$, $x_{1}$, and $x_{2}$. 
Moreover, the translation symmetry is preserved and the solution is given as a function of $x_{1}-x_{2}$. 
Thus, the structure of this solution is determined by three parameters: $\theta_{1}$, $\theta_{2}$, and $x_{1}-x_{2}$. 
The two complex kinks solution (\ref{eq:2kink}) is the solution of the following nonlinear Schr{\"o}dinger equation\,(NLSE).
\begin{equation}
	c_{1} \Delta -i c_{2} \partial_{x}\Delta + c_{3}\left( -\partial_{x}^{2} \Delta + 2 \left|\Delta \right|^{2} \Delta \right) + i c_{4} \left( \partial_{x}^{3}\Delta -6 \left| \Delta \right|^{2} \partial_{x}\Delta \right) =0,
	\label{eq:NLSE1}
\end{equation}
where the parameters $c_{i}$\,$(i=1,2,3,4)$ are related to $m$, $\theta_{1}$, and $\theta_{2}$
in the following manner,
\begin{equation}
	c_{1}=-2 c_{3}m^{2}, \hspace{1em} c_{2}=2m^{2}\left(2\cos\theta_{1}\cos\theta_{2}-1 \right), \hspace{1em} c_{3}= -2m\left(\cos\theta_{1}+\cos\theta_{2} \right).
	\label{eq:coeff}
\end{equation}
Here, we set $c_{4}=1$ without loss of generality.
This type of NLSE is derived from the generalized GL expansion of the grand potential in the vicinity of the tricrtical point\,\cite{Basar2008ki,Thies}. 
The renormalized grand potential density is written as
\begin{equation}
	\Omega_{\rm GL}=c_{0} + c_{1}\left| \Delta \right|^{2} + c_{2} \Im\left( \Delta \partial_{x}\Delta^{*} \right) + c_{3} \left( \left|\Delta \right|^{4}+ \left| \partial_{x}\Delta \right|^{2} \right) + c_{4} \Im \left[ \left(\partial_{x}^{2}\Delta-3 \left|\Delta \right|^{2} \Delta \right)\partial_{x}\Delta^{*} \right] + \cdots.
	\label{eq:potential}
\end{equation}
Here, dots denote higher derivative terms and higher order potential terms. In fact, the NLSE (\ref{eq:NLSE1}) can be derived by considering the variation of the grand potential up to the order of $c_{4}$ with respect to $\Delta^{*}$. 

Now let us consider if the two complex kinks condensate obtained from our calculation agrees with the analytic solution. 
By fitting the analytic solution (\ref{eq:2kink}) with proper scaling to our numerical results, we find that the two complex kinks condensate is well described by\,(\ref{eq:2kink}). Note that $\Delta = \left< O_{2}\right> e^{i \gamma}$ in our convention. 
The fitting results of the amplitude and the phase of order parameter in the two complex kinks condensate are, respectively, shown in the left panel and the right panel of Fig.\,\ref{fig:fit2}. 
The fitting details are given in Appendix\,A.
\begin{figure}[tbp]
	\centering
	\includegraphics[width=7.5cm]{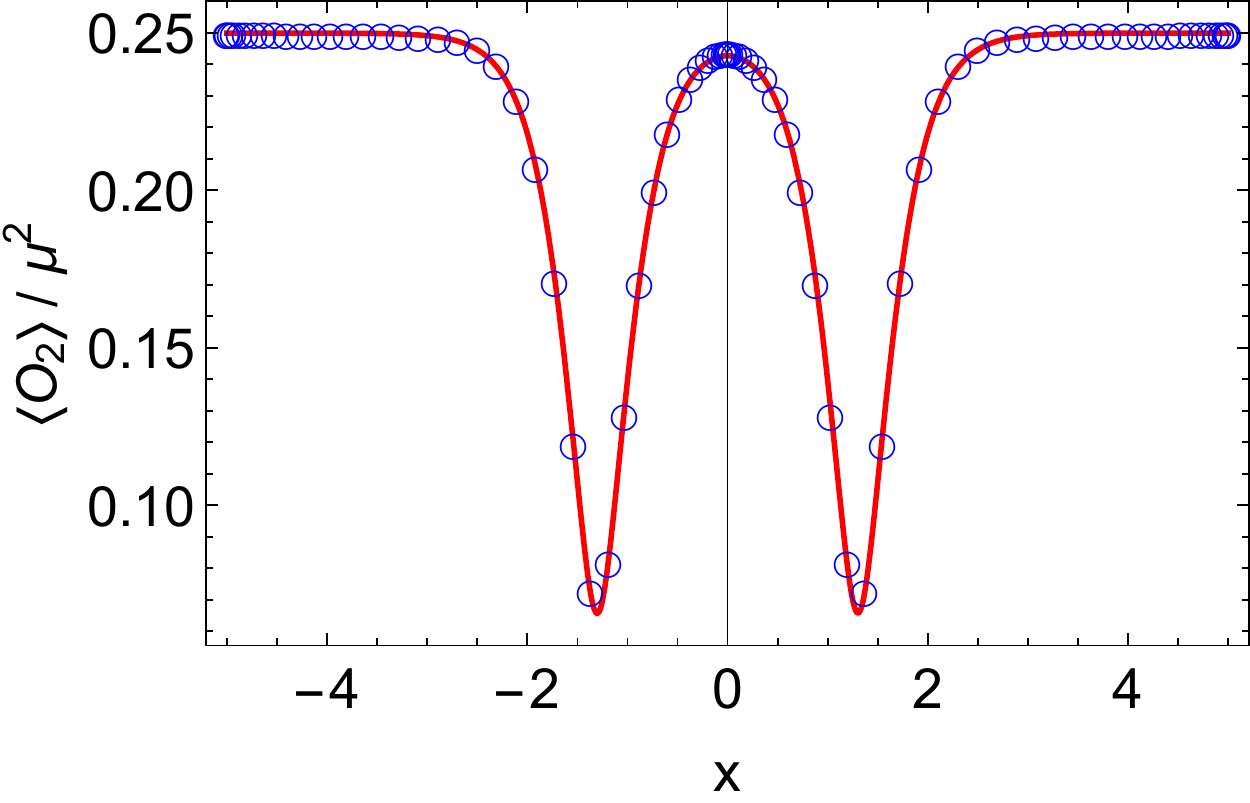}
	\includegraphics[width=7.5cm]{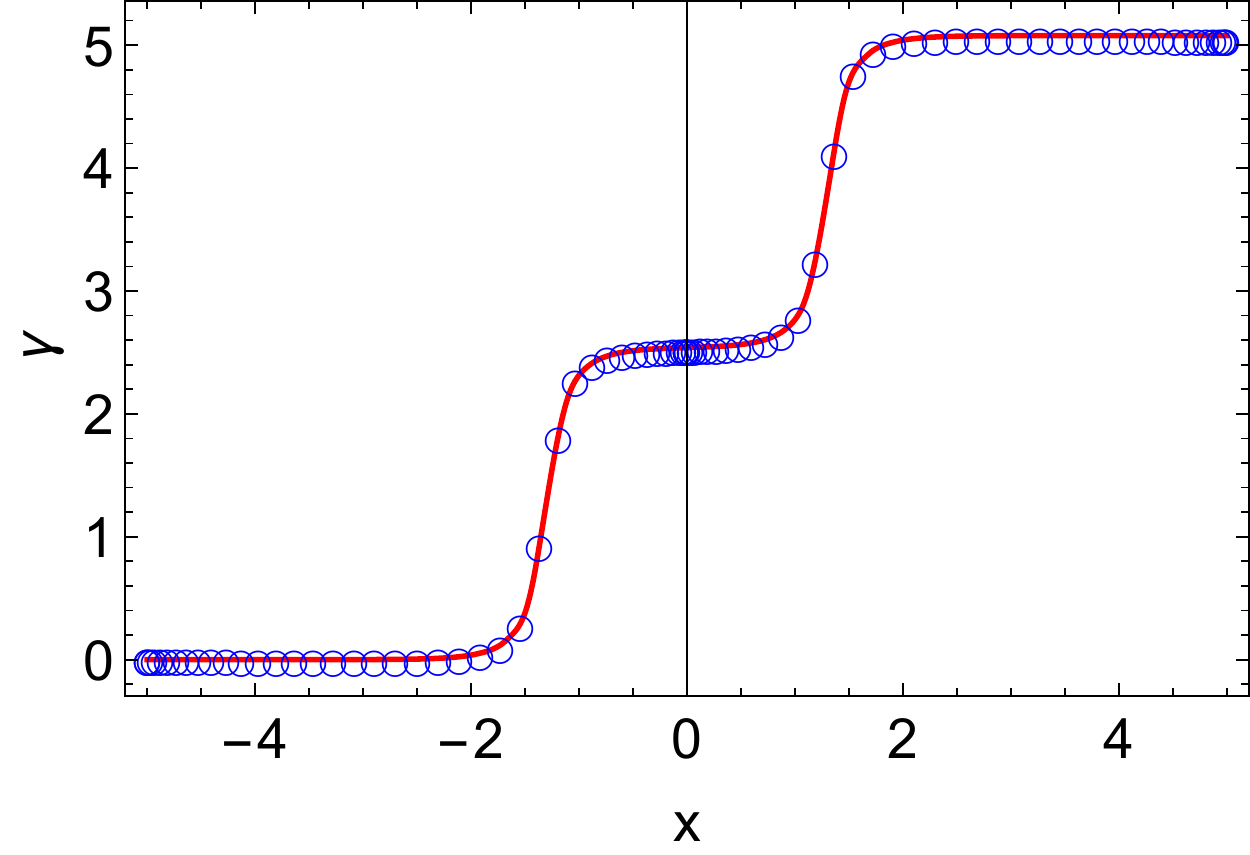}
	\caption{The plots show the amplitude\,(left panel) and the phase\,(right panel) of the order parameter in two complex kinks condensates for $T/\mu=0.0318$ and $J/\mu^{2}=0.0711$, respectively. Open circles and solid line, respectively, denote numerical plots and the analytic function obtained from (\ref{eq:2kink}) with the best fitting parameters.}
	\label{fig:fit2}
\end{figure}
Since the fitting paramters determine the GL paramters $c_{i}$ in\,(\ref{eq:potential}), it quantitatively implies that the boundary theory in the holographic superconductor model can be effectively described by GL theory with higher derivative terms and higher order potential terms. 
In other words, we find that holographic superconductor model represents the boundary physics beyond the conventional GL theory only containing $c_{0}$, $c_{1}$, and $c_{3}$, in which multiple complex kinks solutions cannot be found. 

Furthermore, we also confirm if the twisted kink crystal condensate is well described by the analytic solution derived from the GN model and the NJL model. The twisted kink crystal solution is explicitly given by\,\cite{Basar2008,Basar2008ki,Thies}
\begin{equation}
	\Delta(x) = - A \frac{\sigma\left( Ax + i {\bf K}' - i\theta /2 \right)}{\sigma\left(Ax + i {\bf K}' \right)\sigma\left( i\theta /2 \right)} \exp \left[ iAx\left( -i\zeta(i\theta /2) + i {\rm ns}(i \theta /2) \right) + i\theta \zeta(i {\bf K}') /2 \right],
	\label{eq:TKC}
\end{equation}
where ${\rm ns}$ is the Jacobi elliptic function and the functions $\sigma$ and $\zeta$ are the Weierstrass sigma and zeta functions. 
The real and imaginary half periods are given by $\omega_{1}={\bf K}(\nu)$ and $\omega_{3}=i{\bf K}'\equiv i{\bf K}(1-\nu)$, where ${\bf K}(\nu) = \int_{0}^{\pi/2} = dt / \sqrt{1-\nu \sin^{2}t}$ is the complete elliptic integral. 
The range of the parameter $\theta$ is $0 \leq \theta \leq 4 {\bf K}'(\nu)$. 
The real constant $A$ is a function of $\theta$ and $\nu$: $A(\theta,\nu)=-2i\,{\rm sc}(i\theta/4;\nu){\rm nd}(i\theta/4;\nu)$, where ${\rm sc}$ and ${\rm nd}$ are the Jacobi elliptic functions controlled by the elliptic parameter $\nu$. 
The range of $\nu$ is $0\leq\nu\leq 1$ and the twisted kink crystal is reduced into the single complex kink in the limit of $\nu \rightarrow 1$. 
The amplitude and the phase of order parameter are written as
\begin{eqnarray}
	\left| \Delta(x) \right|^{2} &=& A^{2} \left( {\cal P}(Ax+i{\bf K}') - {\cal P}(i\theta /2) \right), \\
	\gamma(x) &=& A\left( -i\zeta(i\theta /2) + i {\rm ns}(i\theta/2) \right)x +\frac{i}{2} \ln \left( \frac{\sigma(Ax+i{\bf K}'+i\theta /2)}{\sigma(Ax+i{\bf K}'-i\theta /2)} \right) + \frac{\zeta(i {\bf K}')\theta}{2},
\end{eqnarray}
where ${\cal P}$ is the Weierstrass ${\cal P}$ function. 
The twisted kink crystal solution (\ref{eq:TKC}) is the solution of the NLSE (\ref{eq:NLSE1}) with $c_{4}=0$ and this NLSE can be derived from the GL expansion of the grand potential up to the order of $c_{3}$\,\cite{Basar2008ki,Thies}. The coefficients of each term are related to the parameters of the solution,
\begin{equation}
	c_1=A^{2}\left[ 3{\cal P}(i\theta/2)-{\rm ns}^{2}(i\theta/ 2)\right],\hspace{1em} c_2 = -2A\,i\,{\rm ns}\left(i\theta/2 \right).
\end{equation}
Here, we set $c_{3}=1$ without loss of generality. 

In the same way as the two complex kinks condensate, we fit the analytic solution of the twisted kink crystal solution\,(\ref{eq:TKC}) with proper scaling to the our numerical results\,(see Appendix A for details).
Fig.\,\ref{fig:fitTKC} shows that the fitting results of the amplitude and the phase of the order parameter in the twisted kink crystal condensates. 
The fact that our inhomogeneous condensates are well-fitted by the analytic form of the twisted kink crystal solution (\ref{eq:TKC}) again implies that the boundary theory can be represented by the GL theory with higher corrections.
\begin{figure}[tbp]
	\centering
	\includegraphics[width=7.5cm]{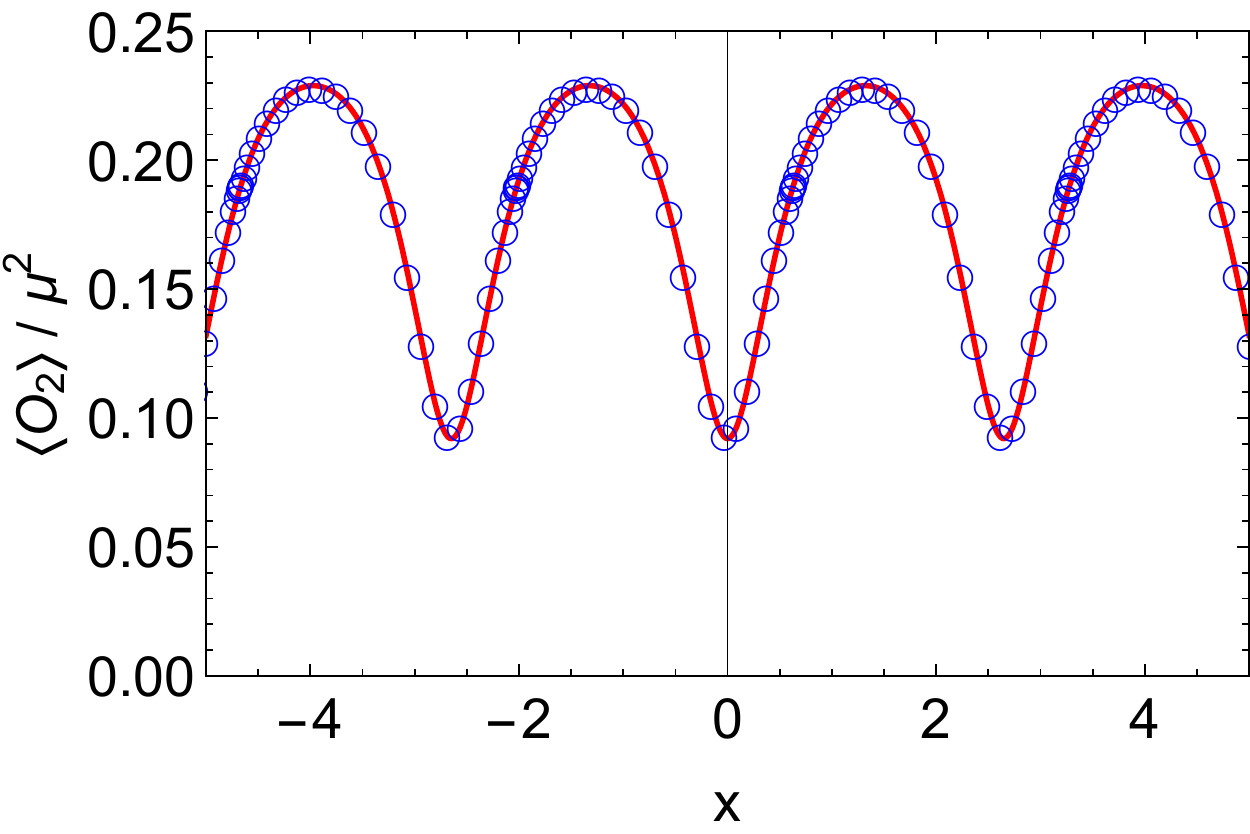}
	\includegraphics[width=7.5cm]{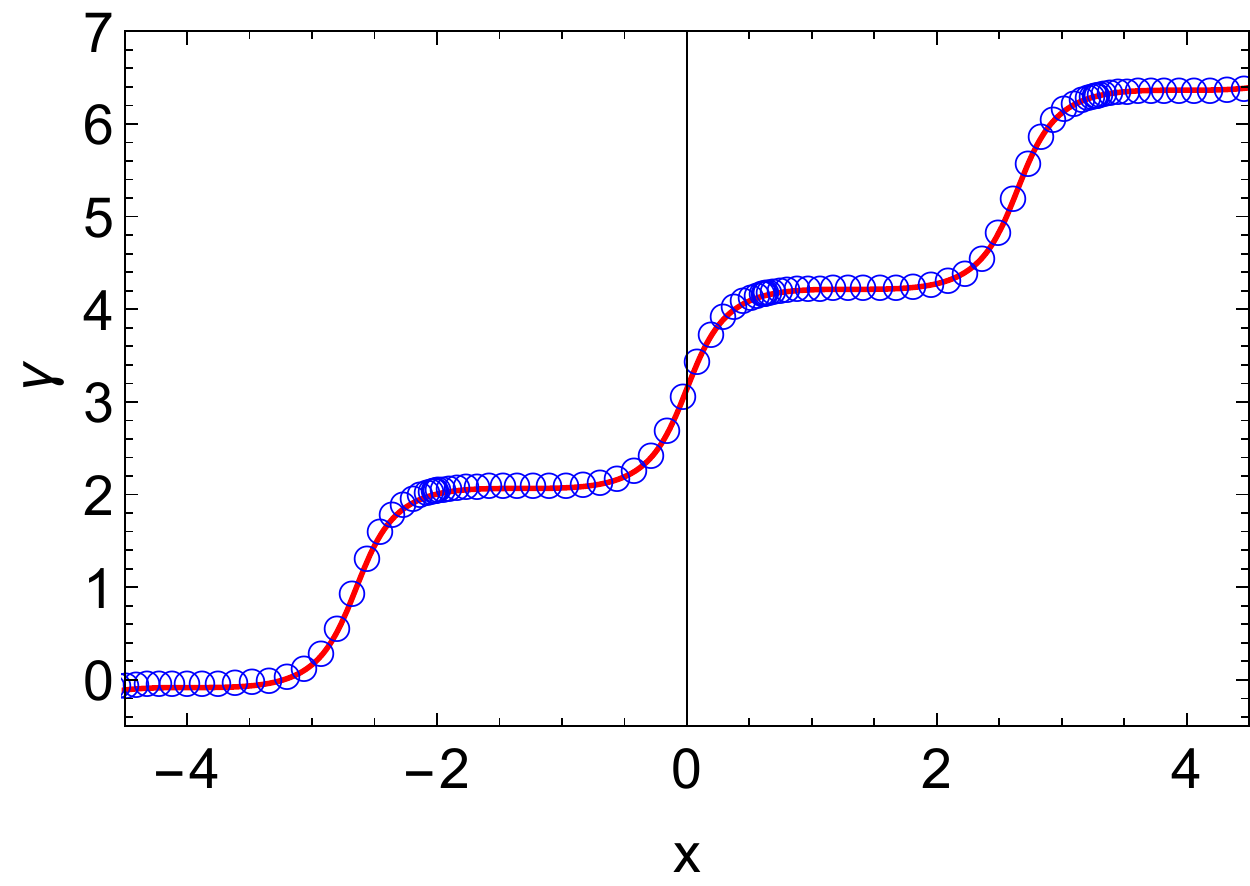}
	\caption{The plots show the amplitude\,(left panel) and the phase\,(right panel) of the order parameter in the twisted kink crystal condensate for $T/\mu=0.0341$ and $J/\mu^{2}=0.327$, respectively. Open circles and solid line, respectively, represent numerical plots and the analytic function obtained from\,(\ref{eq:TKC}) with the best fitting parameters.}
	\label{fig:fitTKC}
\end{figure}

Our results show that the fitting parameters in the analytic solutions determine each coefficient $c_{i}$ in the GL expansion of the grand potential. As a result, we find that the two complex kinks condensates and the twisted kink crystal condensates obtained from our calculations can be effectively described by the GL theory with the grand potential (\ref{eq:potential}). Note that the grand potentials for each solution do not have the same parameters $c_{i}$ because we consider the different order of the GL expansion for them.
Here, we also note that the above analytic 
results from the (1+1) dimensional GN or NJL model are derived in the vicinity of tricritical point, although our boundary theory is in (2+1) dimensional spacetime at an arbitrary finite temperature. 
Also, the infinite spatial region is considered in the analytic approach whereas we have to consider a sufficiently large but finite spatial region for numerical computations. 
We consider that these differences give rise to a slight modification for fitting such as a proper scaling\,(Appendix\,A). 
Nevertheless, we find that the inhomogeneous condensates we obtained in the holographic superconductor model can be well described by those analytic forms. 
Therefore, our results strongly support the expectation that the holographic superconductor model represents the GL theory with higher corrections beyond the conventional GL theory.

\subsection{Free energy}
In this section, we compute the free energy of the inhomogeneous solutions.
The free energy of the dual field theory is defined from the on-shell bulk action $\Omega=-TS_{\rm os}$.
The on-shell bulk action in our setup is explicitly given by
\begin{equation}
\begin{split}
	S_{\rm os} = -\int d^{3}x  \left. \left[\frac{1}{2} M_{t} M_{t}'  +\frac{f}{2}(\partial_{x}M_{z} -M_{x}')M_{x} -\frac{f}{z^2}\psi\psi' \right] \right |_{z=0} \\
-  \int d^{4}x \frac{\psi^{2}}{z^2}\left( \frac{M_{t}^2}{f}-f M_{z}^2-M_{x}^2 \right),
\end{split}
\end{equation}
where the first term is evaluated at the AdS boundary $z=0$.
For our inhomogeneous solutions, the free energy can be explicitly written as
\begin{equation}
	\frac{\Omega}{\rm Vol} = \frac{1}{l} \int^{l/2}_{-l/2} dx  \left[ \frac{1}{2}\Bigl( \mu \rho(x) - \nu(x) J  \Bigr) + \frac{1}{2}  \int^{1}_{0} dz  \left( \frac{M_{t}^{2}}{f} - M_{x}^{2} - f M_{z}^{2} \right)\phi^{2} \right],
\end{equation}
where ${\rm Vol} = \int dz dx dy =l \int dy$ in our setup.
We evaluate the thermodynamic stability of the inhomogeneous solutions by defining $\Delta \Omega $ as the difference of the free energy between the superconducting state and the normal state with the corresponding value of $T/\mu$. 
Note that we cannot compare the free energy of the superconducting state with that of the normal state with the same values of $T/\mu$ and $J/\mu^{2}$ since it is difficult to prepare the normal state with a finite current (a similar issue on the shortcoming of this method is also mentioned in\,\cite{Arean2010}).

\begin{figure}[tbp]
	\centering
	\includegraphics[width=14cm]{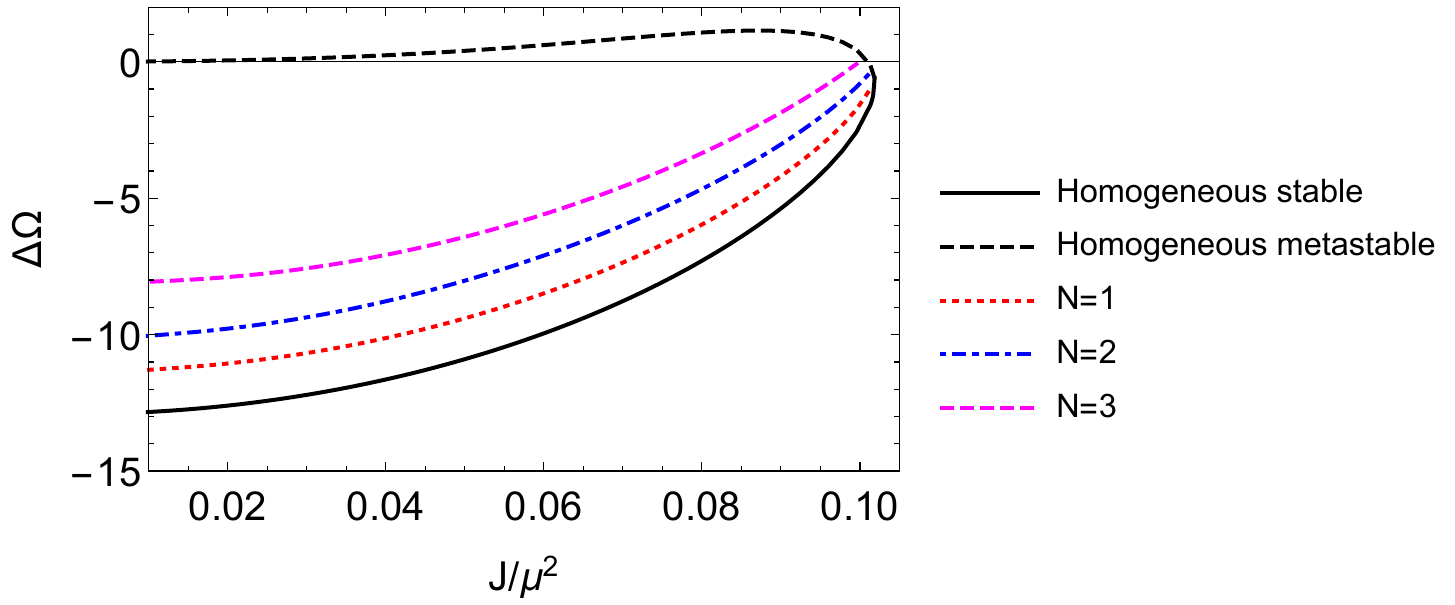}
	\caption{The difference of the free energy between the superconducting state and the normal state as a function of $J/\mu^{2}$ with $T/\mu=0.0318$. Each plot represents the free energy in the solutions with different numbers of kinks.}
	\label{fig:free}
\end{figure}

In Fig.\,\ref{fig:free}, we plot $\Delta \Omega$ as a function of $J/\mu^{2}$ with $T/\mu$ fixed.
Here, we show $\Delta\Omega$ for different solutions:\,a stable homogeneous condensate, a metastable homogeneous condensate, and a complex kink(s) condensate. 
In Fig.\,\ref{fig:free}, $N$ stands for the numbers of kinks in the solutions.
As studied in\,\cite{Arean2010}, there are two different homogenous condensates in the presence of the constant current. It has been shown that one is thermodynamically stable and the other is metastable by computing the free energy.
Comparing the value of $\Delta\Omega$ for the complex kink(s) condensate to those for the homogeneous condensates, we find that the complex kink(s) condensate is metastable for any value of $J/\mu^2$. 
Also, we find that the free energy becomes larger for more kinks and these solutions are also metastable.
Our result is consistent with the GN theory or the BCS theory for low density at low temperatures, since the inhomogeneous solutions such as the complex kink(s) condensate or the twisted kink crystal condensate appear and they are metastable in these theories. 
This fact is important due to the following reason. The existence of the metastable (not unstable) inhomogeneous condensates not only agrees with the microscopic theory such as the GN theory or BCS theory, but also implies that these inhomogeneous condensates could be stable in another setup. In fact, the LOFF phase is realized in the presence of a large magnetic field. Therefore, our findings of the inhomogeneous condensates are a preliminary step towards realizing the one-dimensional inhomogeneous condensates in the minimum setup of the holographic superconductor.

\section{Conclusion and Discussion} \label{sec:4}
\begin{figure}[tbp]
	\centering
	\includegraphics[width=10cm]{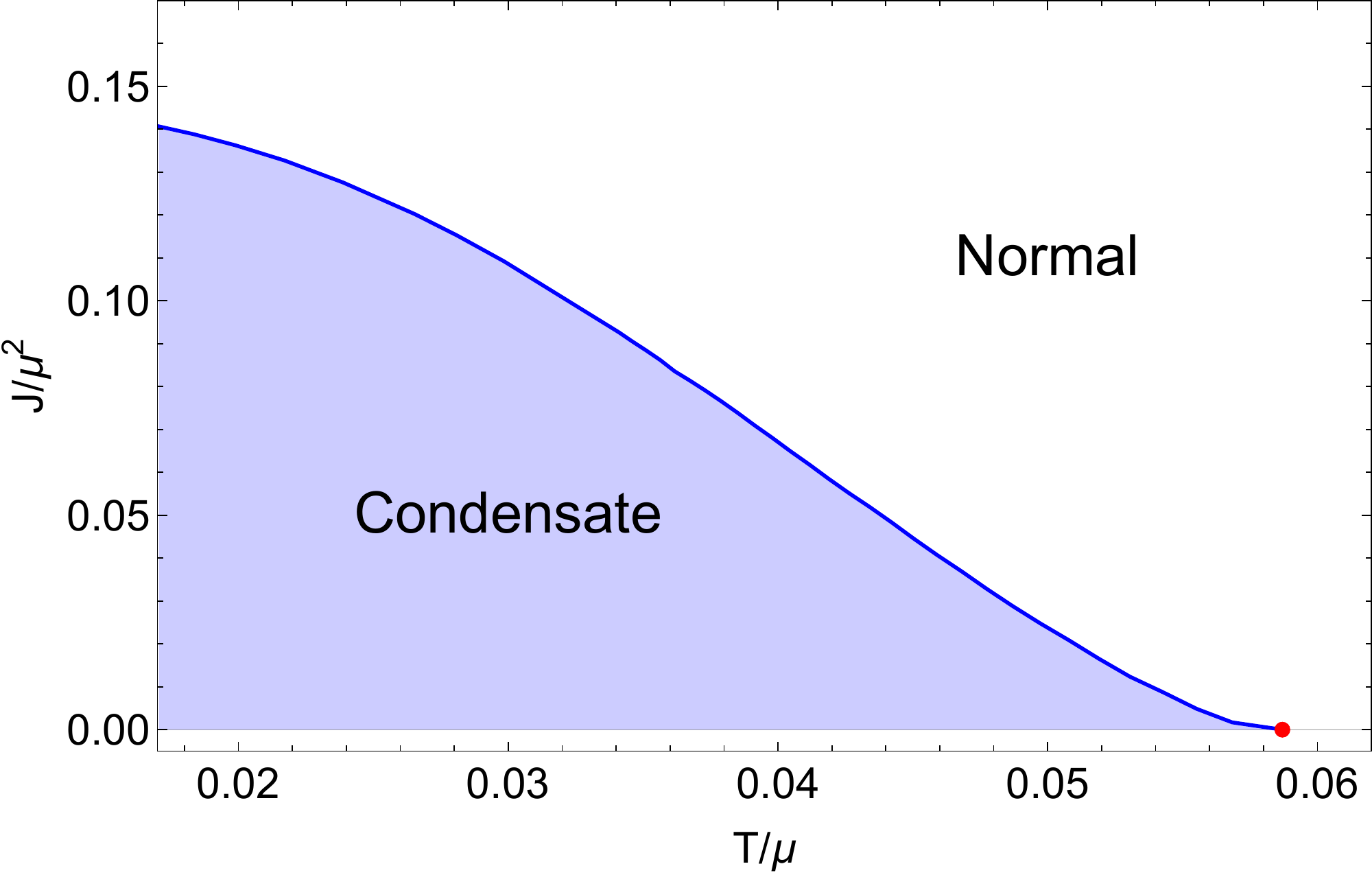}
	\caption{The phase diagram of the holographic superconductor model with a constant current. The red dot corresponds to the criticial temperature $T/\mu \approx 0.0587$. In the condensate phase, we find not only the homogeneous condensate but also the single complex kink condensate, the multiple complex kinks condensate, and the twisted kink crystal condensate. The phase boundary of the homogeneous condensate denoted by the blue curve is coincident with that of the complex kink(s) condensates.}
	\label{fig:phase}
\end{figure}

In this paper, we study the inhomogeneous solutions which appear in the (3+1)-dimensional holographic superconductor model.
In our previous work\,\cite{Matsumoto2019}, we showed that the holographic superconductor model reproduces the inhomogeneous solutions which have the spatially inhomogeneous amplitude of the order parameter without a current. 
In this work, adding to these solutions, we also show the different inhomogeneous solutions, the single complex kink condensate, the multiple complex kinks condensate, and the twisted kink crystal condensate, by numerical calculations in the framework of the holographic superconductor model.
In these solutions, both the amplitude and the phase of the order parameter spatially modulate in the presence of the current.
We find that the amplitude of the order parameter for these solutions is finite at the position of the kink, whereas it is zero for the real kink solutions and the real multi-kink solutions\,\cite{Matsumoto2019}.
This behaviour is consistent with the behaviour of the inhomogeneous solution found in the previous investigations based on the BCS, NJL, or GN models for low density\,\cite{Thies, Nickel2009, Buballa2015}.
In Fig.\,\ref{fig:phase}, we show the phase diagram of the holographic superconductor model with a constant current. In the condensate phase, one can find not only the homogeneous condensate but also the single complex kink condensate, the multiple complex kinks condensate, and the twisted kink crystal condensate. We find that the phase boundary of the homogeneous condensate denoted by the blue curve in Fig.\,\ref{fig:phase} is coincident with that of the complex kink(s) condensates. In other words,  both the homogeneous condensates and the complex kink(s) condensates can be found in the condensate phase.

It is constructive to compare our results with the moving kink solution at zero temperature\,\cite{Gao2019}. 
Our solutions correspond to the static inhomogeneous solutions in which the kink structure does not have a relative velocity with respect to the thermal bath. 
On the other hand, for the moving kink solution, the superconducting component has a relative velocity with respect to the normal component\,(namely the thermal bath), since the non-zero super current is present.
Since mutual friction is absent in our static solutions, it would imply that the mutual friction could appear only when there is a relative velocity between the kink structure and the normal component. 

We also study the current dependence of the total gauge-invariant phase difference in the single complex kink condensate. 
It is likely to approach $2\pi$ in the limit of the zero current and goes to zero for a larger current. 
At high temperatures, the total gauge-invariant phase difference is a monotonically decreasing function of the current. 
At low temperatures, on the other hand, we find that it becomes non-monotonic with respect to the current. 
This result implies the existence of the phase transition at specific temperatures.

We also study the boundary interpretation of those inhomogeneous condensates by comparing the analytic solutions derived from the (1+1) dimensional GN model and NJL model. 
We perform the fitting of them to the two complex kinks condensate and the twisted kink crystal condensate. 
As a result, we find that they are well-fitted by the known analytic solutions. 
Since these analytic solutions satisfy the NLSE which is derived from GL theory with higher corrections such as the higher derivative terms and the higher order potential terms, our results imply that the boundary theory can also be effectively described by the GL theory with higher corrections. 
In other words, it is expected that holographic superconductor model represents the boundary physics beyond the conventional GL theory.

Moreover, we compute the free energy of the complex kink(s) condensate as a function of the current and compare it to those of the homogeneous solutions.
We find that the free energy of the complex kink(s) condensate is always higher than that of the homogeneous solution for given $J/\mu^2$.
Our result implies that the complex kink(s) condensates are metastable.
In the microscopic theories such as GN theory or BCS theory, it is known that the complex kink(s) condensates and the twisted kink crystal condensate are metastable.
Thus, our calculation of the free energy agrees with the implications in the previous studies.
Though it is known that the inhomogeneous phase becomes stable for a larger chemical potential\,\cite{Thies, Nickel2009, Buballa2015},
the behaviour is not observed in the present study.
The region in which the inhomogeneous phase becomes stable would be approachable by improving the numerics; however, it is beyond the scope of this paper.

Before ending the conclusion, we have few remarks.
In our study, we ignore the backreaction to the matter sector from the gravity sector by taking the probe limit. 
Since the backreaction should be important at low temperatures, it would be interesting to consider the effect of the backreaction on the non-monotonic behaviour we found.
We also ignore the magnetic field which is known to stabilize the inhomogeneous solutions due to the Zeeman effect and the Aharonov-Bohm effect \cite{Yoshii2015}.
It would be interesting to consider the spin degrees of freedom and the magnetic field. 
Also, it is straightforward to extend our study to the $p$-wave superconductor\,\cite{Gubser2008wv,Wang2011}, although we focus on the $s$-wave superconductor in this paper.
Along this direction, it would be interesting to study the competition between the $s$-wave and the $p$-wave superconductor\,\cite{Liu2015} for the inhomogeneous condensate we found. 
Recently, the nonequilibrium process of the inhomogeneous condensate was studied in\,\cite{Guo2018}. 
The dynamical behaviour of the inhomogeneous solutions we found in this paper should be investigated. 
In another direction, it is also interesting to investigate the hydrodynamic behaviour of the Nambu-Goldstone mode in our case by studying the quasi-normal mode in holographic superconductor\,\cite{Amado2009}.\footnote{This analysis could be extended to the nonequilibrium regime as studied in Ref.\,\cite{Ishigaki2020}, although it focuses on the spontaneous chiral symmetry breaking.}
For the gravity side, the interpretation of the (meta-)stable inhomogeneous solutions are still lacking.
We leave them for future works.

\section*{Acknowledgment}
We thank Shunichiro Kinoshita, Shin Nakamura, and Muneto Nitta for helpful discussions. 
We are also grateful to Yu Tian for fluitful comments.
R.Y.\,is especially grateful to Hikaru Watanabe and Asahi Yamaguchi for fruitful discussions and comments.
The work of M.M.\,is supported by National Natural Science Foundation of China Grant No.\,12047538.
The work of R.Y.\,is supported by JSPS Grant-in-Aid for Scientific Research (KAKENHI Grants No. 19K14616 and No. 20H01838).

\section*{Appendix A: Fitting details}
In this appendix, we show the details of fitting performed in section III D. The analytic solution of the two complex kinks condensate (\ref{eq:2kink}) is derived under the assumption of the infinite range of the spatial direction $x$. On the other hand, we fix the spatial range $-l/2 \leq x \leq \l/2$ for numerical calculations. Then, we perform the rescaling $x\rightarrow a x$ in\,(\ref{eq:2kink}) and consider $a$ as one of parameters. Also, for convenience, in order to perform the fitting of the phase, we fit the $x$ derivative of the phase in\,(\ref{eq:2kink}) to the velocity $\nu(x)$. The best fitting parameters in Fig.\,\ref{fig:fit2} are $m=0.249873$, $\theta_{1}=1.28987$, $\theta_{2}=1.31977$, $x_1=x_2=-2.16001$, and $a=1.93453$ for the amplitude\,(left panel) and  $m=0.249873$, $\theta_{1}=1.26635$, $\theta_{2}=1.29073$, $x_1=x_2=-2.27880$, and $a=2.03423$ for the phase\,(right panel). Substituting the fitting parameters ($m,\theta_{1},\theta_{2}$) into Eq.\,(\ref{eq:coeff}), the corresponding GL parameters are determined.

In the fitting for the twisted kink crystal condensate, the form of\,(\ref{eq:TKC}) is not well-fitted to our numerical results. Alternatively, we assume the following form:
\begin{equation}
	\Delta(x) = - B \frac{\sigma\left( Cx + i {\bf K}' - i\theta /2 \right)}{\sigma\left(Cx + i {\bf K}' \right)\sigma\left( i\theta /2 \right)} \exp \left[ iCx\left( -i\zeta(i\theta /2) + i {\rm ns}(i \theta /2) \right) + i\theta \zeta(i {\bf K}') /2 \right],
	\label{eq:TKC2}
\end{equation}
where the parameters $B$ and $C$ are independent of each other. Using this form, we obtain the well-fitted results as shown in Fig.\,\ref{fig:fitTKC}. The best fitting parameters in Fig.\,\ref{fig:fitTKC} are $B=0.242465$, $C=1.62465$, $\theta=2.54102$, and $\nu=0.74711$ for the amplitude\,(left panel) and $C=1.38987$, $\theta=2.98549$, and $\nu=0.481891$ for the phase\,(right panel). We consider that the reason why we have to use the alternative form (\ref{eq:TKC2}) can be related to the difference of the spacetime dimension, the finite spatial range, and so on, as discussed in the main text. Note that one can confirm that the alternative form of the twisted kink crystal solution (\ref{eq:TKC2}) is also the solution to NLSE (\ref{eq:NLSE1}) with $c_{4}=0$. That is, one can check that the analytic form (\ref{eq:TKC2}) satisfies the following NLSE,
\begin{equation}
	\partial_{x}^{2}\Delta - 2 \left|\Delta\right|^{2} \Delta - \left[ 2\left(C^{2}-B^{2}\right){\cal P}(Cx+ i {\bf K}') + \left(C^{2}+2B^{2} \right){\cal P}(i\theta /2) -C^{2}{\rm ns}^{2}(i\theta/2) \right]\Delta -2iC\left[i{\rm ns}(i\theta/2) \right]\partial_{x}\Delta=0.
\end{equation}
This implies that the alternative form of the twisted kink crystal solution (\ref{eq:TKC2}) can also be derived from GL theory with higher corrections. In other words, the four parameters $(B,C,\theta,\nu)$ determine the GL paramters $c_{i}$ according to (\ref{eq:NLSE1}) with $c_{4}=0$ and (\ref{eq:potential}) via the following relation
\begin{equation}
\begin{aligned}
	c_{1} &=  \left[ 2\left(C^{2}-B^{2}\right){\cal P}(Cx+ i {\bf K}') + \left(C^{2}+2B^{2} \right){\cal P}(i\theta /2) -C^{2}{\rm ns}^{2}(i\theta/2) \right], \\
	c_{2} &= -2C\,i\,{\rm ns}(i\theta/2), \hspace{1em}c_{3}=1.
\end{aligned}
\end{equation}

\end{document}